\documentclass[twocolumn,prb,nofootinbib]{revtex4-2}
\usepackage{graphicx}
\usepackage{color}
\usepackage{bm}
\usepackage{upgreek}
\usepackage{comment}
\usepackage{ulem}   

\begin{document}
\newcommand{\RED}{\color{red}}
\newcommand{\BLUE}{\color{blue}}

\newcommand{\taum}{\uptau_{\rm m}}

\title{Emergent topological quasiparticle kinetics in constricted nanomagnets}

\author{J. Guo$^{1,\dagger}$}
\author{D. Hill$^{1,\dagger}$}
\author{V. Lauter$^{2}$}
\author{L. Stingaciu$^{2}$}
\author{P. Zolnierczuk$^{2}$}
\author{C. A. Ullrich$^{1,*}$}
\author{D. K. Singh$^{1,*}$}

\affiliation{$^{1}$Department of Physics and Astronomy, University of Missouri, Columbia, MO}
\affiliation{$^{2}$Oak Ridge National Laboratory, Oak Ridge, TN 37831, USA}
\affiliation{$^\dagger$These authors contributed equally: J. Guo and D. Hill}
\affiliation{$^*$Email: ullrichc@missouri.edu, singhdk@missouri.edu}

\maketitle

\textbf{The ubiquitous domain wall kinetics under magnetic field or current application describes the dynamic properties in nanostructured magnets.\cite{Atkinson,Coey,Klaui} However, when the geometrical size of a nanomagnetic system is constricted to the limiting domain wall length scale, the competing energetics between anisotropy, exchange and dipolar interactions can cause emergent kinetics due to quasiparticle relaxation, similar to bulk magnets of atomic origin.\cite{Bigot,Braun} Here, we present a joint experimental and theoretical study to support this argument -- constricted nanomagnets, made of antiferromagnetic and paramagnetic neodymium thin film with honeycomb motif, reveal fast kinetic events at ps time scales due to the relaxation of chiral vortex loop-shaped topological quasiparticles that persist to low temperature in the absence of any external stimuli. Such phenomena are typically found in macroscopic magnetic materials. Our discovery is especially important considering the fact that paramagnets or antiferromagnets have no net magnetization. Yet, the kinetics in neodymium nano\-structures is quantitatively similar to that found in ferromagnetic counterparts and only varies with the thickness of the specimen. This suggests that a universal, topological quasiparticle mediated dynamical behavior can be prevalent in nanoscopic magnets, irrespective of the nature of underlying magnetic material.}

Advances in nanomagnetism \cite{Cowburn,Bader} have led to a paradigm shift in technologies that harness magnetism dynamics on a nanoscale, such as magnetic memory devices, magnetic tunnel junctions for spintronics applications and spin torque nano-oscillators as miniaturized microwave sources for electronic circuits.\cite{Chien,Boona,Ralph,Urazhdin,Bhowmik} Conventionally, domain wall kinetics is considered to be the driving mechanism behind the dynamic behavior in nanostructured magnets, which requires magnetic field or electric current application.\cite{Coey,Shibata} However, at length scales approaching the single domain limit, which defines the so-called constricted nanomagnets, the nature of magnetic interactions and the ensuing dynamic properties change dramatically. At such small geometrical length scales, the competing energetics between exchange, anisotropy and dipolar terms causes inherent fluctuation in the macroscopic magnetic correlation parameter $\langle {\bf m}_i \cdot {\bf m}_j\rangle$.\cite{Braun} Consequently, a new dynamic mechanism can emerge, as evidenced by the recent numerical simulations of constricted nanomagnets using the Landau-Lifshitz magnetization model.\cite{Zhou} Here we test this hypothesis in nanoscopic honeycomb shaped neodynium (Nd) lattices with narrow structural components, using neutron scattering combined with micromagnetic simulations. We chose Nd for this study because it acts as an antiferromagnet (AFM) below $T_N < $25 K and behaves as a paramagnet (PM) above $T_N$.\cite{Bak} As shown schematically in \textbf{Fig. 1} and explained in more detail below, vortex loop-shaped quasiparticles arise in the Nd nanostructures in the absence of any magnetic field. A similar behavior is found in ferromagnetic (FM) systems of the same geometry, but the vortex loop size is bigger in the FM lattice compared to AFM or PM. Surprisingly, the quasiparticles relax at a very high rate of picoseconds in all three cases, as confirmed by experiment and theory.

\begin{figure*}
\includegraphics[width=\linewidth]{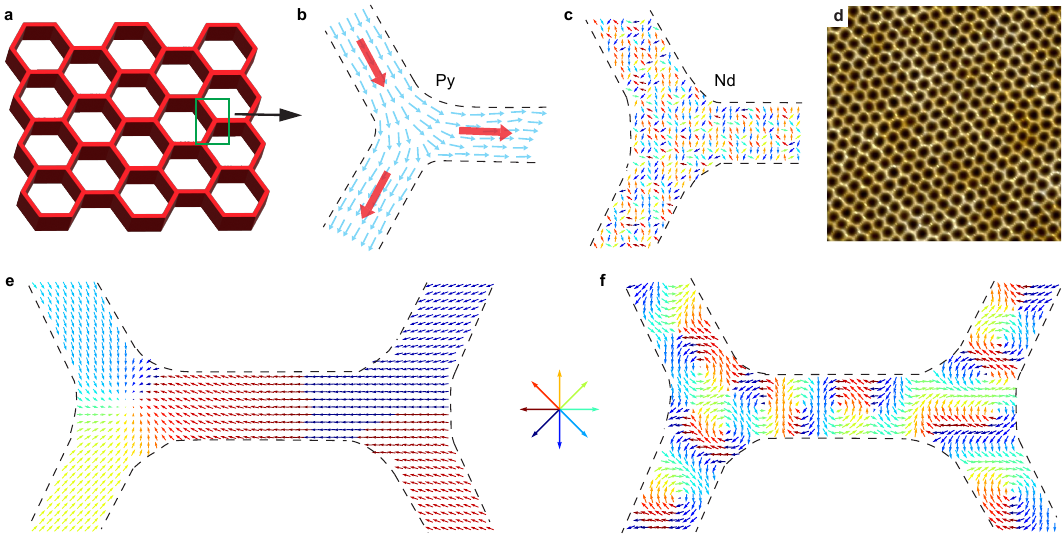}
\caption{\textbf{Quasiparticle mediated dynamics in antiferromagnetic and ferromagnetic honeycomb lattices of constricted nanoscopic elements.} \textbf{a}, Schematic of a typical honeycomb lattice. \textbf{b-c}, Microscopic profile of spin distribution in permalloy and Nd honeycomb element and vertex. While net macroscopic moment develops along the length in permalloy (Py) element due to shape anisotropy, antiferromagnetic and paramagnetic character of Nd forbids moment formation along the constricted honeycomb element (Fig. \textbf{c}). \textbf{d}, Atomic force micrograph of nanoscopic honeycomb lattice, used in this study. \textbf{e-f}, Numerical simulation reveals the existence of chiral vortex loop-shaped quasiparticles in constricted nanomagnet of honeycomb lattice that relax at very fast rate in the absence of any external tuning agent, such as magnetic field or electric current. The chirality is more pronounced in Nd (Fig. \textbf{f}) where the average size of quasi-particle is $\sim$ 3.7 nm. The size of quasi-particle is larger in Py honeycomb (Fig. \textbf{e}). However, in both cases, the magnetic cores of vortex-type topological defects point perpendicular to the plane of the lattice.}
\label{fig1}
\end{figure*}

Magnetic honeycomb lattices have been subject of intense experimental and theoretical investigations for their ability to manifest emergent phenomena due to chiral magnetic correlations and topological magnetic charges.\cite{SH,Peter,Guo2,Yue,Roug} Magnetic charges arise on the honeycomb vertices under the dumbbell prescription of magnetic moments.\cite{Sondhi,Oleg} The magnetic moments are locally arranged in two basic configurations: two-in and one-out or vice versa, and all-in or all-out. Subsequently, net charges of $\pm Q$ and $\pm 3Q$ are imparted to honeycomb vertices, respectively.\cite{SH,Peter2,Tanaka,Ladak,Artur} The interchangeable physical descriptions in a honeycomb lattice based on magnetic moment and magnetic charge concepts render an archetypal platform to comprehensively study the nanoscopic mechanisms behind the underlying dynamics.

It is tempting to assume that an artificially created nanostructured honeycomb lattice would exhibit similar kinetic events at the nanoscale as spin ice magnets at angstrom length scales, the former being a two-dimensional (2D) analogue of the latter. The dynamic phenomena in a spin ice magnet are dictated by the kinetics of gauge invariant effective magnetic monopoles (also termed magnetic charge defects).\cite{Bramwell,Moessner2} Recent studies indeed found evidence of magnetic charge defect kinetics in artificial 2D lattices, albeit in the field-induced environment.\cite{Mengotti,Bhat,Shen,Heyderman,Zeissler,Mellado} More recently, we found evidence of self-propelled magnetic charge defect dynamics in the absence of any external stimuli, in nanoscopic honeycomb lattices made of ultra-small single domain size permalloy (Ni$_{0.81}$Fe$_{0.19}$) elements with connected topography.\cite{Guo} The nanoscale element, with typical size of $11 \times 4 \times 6$ nm
($\mbox{length}\times \mbox{width} \times \mbox{thickness}$), produces small enough inter-elemental dipolar interaction energy ($\sim$ 44 K) and barrier crossing energy for magnetization reversal ($\sim$ 70 K) that make it possible to probe the temperature dependent magnetic charge defect dynamic properties.

However, if the nanostructure is made of a PM or AFM material such as Nd, the dumbbell description does not apply since there are no finite moments along the length of honeycomb element, see the schematic {\bf Fig. \ref{fig1}b}. While the net magnetic moment is zero for an AFM, the weak exchange energy in a PM is insufficient to form a macroscopic moment to enforce the spin ice rule. Therefore, the description of magnetic charge relaxation is not applicable in nanostructured honeycombs made of these materials. Rather, we find that a more exciting dynamic property emerges in this case.

\begin{figure*}
\includegraphics[width=\linewidth]{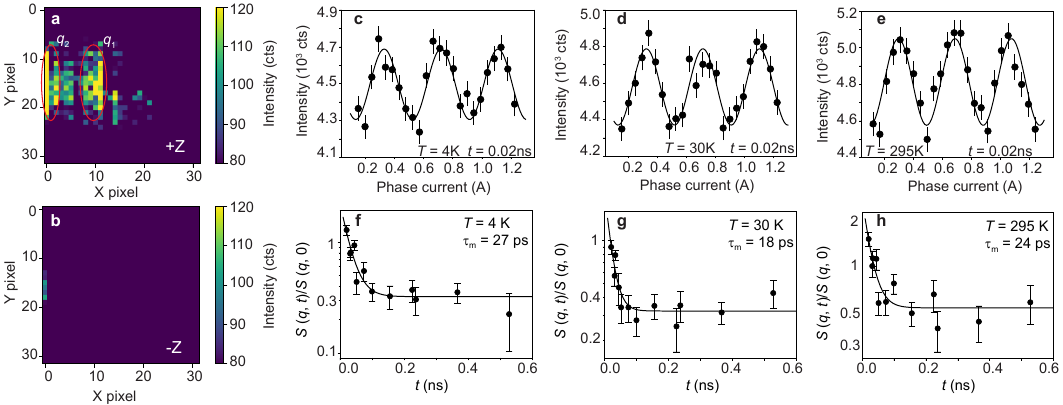}
\caption{\textbf{Magnetic dynamics in the Nd honeycomb nanomagnet.} \textbf{a, b}, neutron spin echo (NSE) spectroscopy reveals localized spectral weight, indicating the presence of dynamic behavior in the system without any external stimuli, such as magnetic field application. Each pixel in the color plots corresponds to a distinguishable wavevector $q$. We observe strong intensity at $q_1 = 0.058$ \AA$^{-1}$~and $q_2 =0.029$ \AA$^{-1}$~for neutron polarization along the +Z axis. The absence of bright intensity for neutron polarization along the -Z axis 
indicates the magnetic nature of the scattering. There is also faint scattering around X pixel 18, corresponding to $q_1 \sim $ 0.08 \AA$^{-1}$. \textbf{c-e}, strong signal-to-background ratio is observed in representative spin echo oscillations at different temperatures at Fourier time $t = 0.02$ ns at $q_1 =  0.059$ {\AA}$^{-1}$. Similar echo profiles are detected at $q_2 =  0.029$ {\AA}$^{-1}$. Error bars represent one standard deviation. \textbf{f-h}, normalized NSE spectral function $S(q, t)/S(q, 0)$ versus neutron Fourier time at $T= 4$ K, 30 K and 295 K at $q_1 =  0.059$ {\AA}$^{-1}$. The data is fitted with exponential functions to extract $\taum$. Error bars are calculated from the square root of the variance of the least-square-fitted parameter.}
\label{fig2}
\end{figure*}

We have fabricated nanoscopic honeycomb lattices of Nd with similar elemental characteristics as the FM structures studied recently \cite{Guo} (see Methods). The atomic force micrograph, shown in \textbf{ Fig. \ref{fig1}f}, confirms the high structural quality of the 2D lattice sample. We investigated the dynamic properties of the Nd nanostructures using neutron spin echo (NSE) spectroscopy (see Methods for details).\cite{NSE2} NSE is a quasi-elastic measurement technique where the relaxation of a magnetic specimen is decoded by measuring the relative change in polarization of scattered neutrons via the change in the phase current at a given Fourier time (related to the precession of neutrons).\cite{NSE2} In \textbf{Fig. \ref{fig2}} we show NSE results at several characteristic temperatures and neutron Fourier times obtained on the Nd honeycomb. The NSE data for spin-up neutron polarization reveal prominent localized soft excitations at $q$ values of $0.058$ {\AA}$^{-1}$ and $0.029$ {\AA}$^{-1}$, see \textbf{Fig. \ref{fig2}a}. Discrete faint scattering is also detected at a higher $q$ value of 0.08 {\AA}$^{-1}$. At the same time, little or no spectral weight is detected in spin-down neutron polarization data, see \textbf{Fig. \ref{fig2}b}, confirming the magnetic nature of the signal (also see Figures S1 and S3 in Supplementary Materials). Quantitative plots of the spin echo profile at prominent $q$ wave vectors depict high quality sinusoidal oscillations, with strong signal-to-background ratio, throughout the measurement temperature range. \textbf{Figs. \ref{fig2}c-e} show spin echo profiles at three characteristic temperatures at Fourier time $t = 0.02$ ns. The reciprocal lattice vectors of $q = 0.058$ {\AA}$^{-1}$ and 0.029 {\AA}$^{-1}$ in \textbf{Fig. \ref{fig2}a} correspond to the distances between the nearest neighboring and the next nearest neighboring vertices, respectively. Thus, the experimental data suggests relaxation phenomena due to kinetic events of magnetic entities between the nearest neighboring and the next nearest neighboring vertices, respectively. Weak spectral weight at higher $q$ indicates that the kinetic process also occurs over a partial length in a honeycomb element and that the probability of such events is smaller than the full element-wide relaxation.

The scattering pattern in \textbf{Fig. 2a} is highly analogous to the NSE data obtained on permalloy honeycomb lattices, which was ascribed to the relaxation of magnetic charge defects between neighboring vertices.\cite{Guo} In the permalloy lattice, magnetic charges on the vertices can undergo transformations by releasing or absorbing charge defects (or monopoles) of magnitude $\pm2Q$  (the difference between the absolute value of low and high multiplicity charges) that traverse the lattice.\cite{Shen,Mengotti} The presence of $\pm 3Q$ charges catalyzes the kinetic process as they are not energetically stable. By contrast, Nd elements have no net magnetic moment, and there can be no dynamics induced by $2Q$ charge defect kinetics. This leads us to conclude that the magnetic relaxation process in Nd honeycomb lattices must have a different origin, or that the magnetic charge defect is more complex than originally assumed and can have universal origin.

\begin{figure*}
\includegraphics[width=\linewidth]{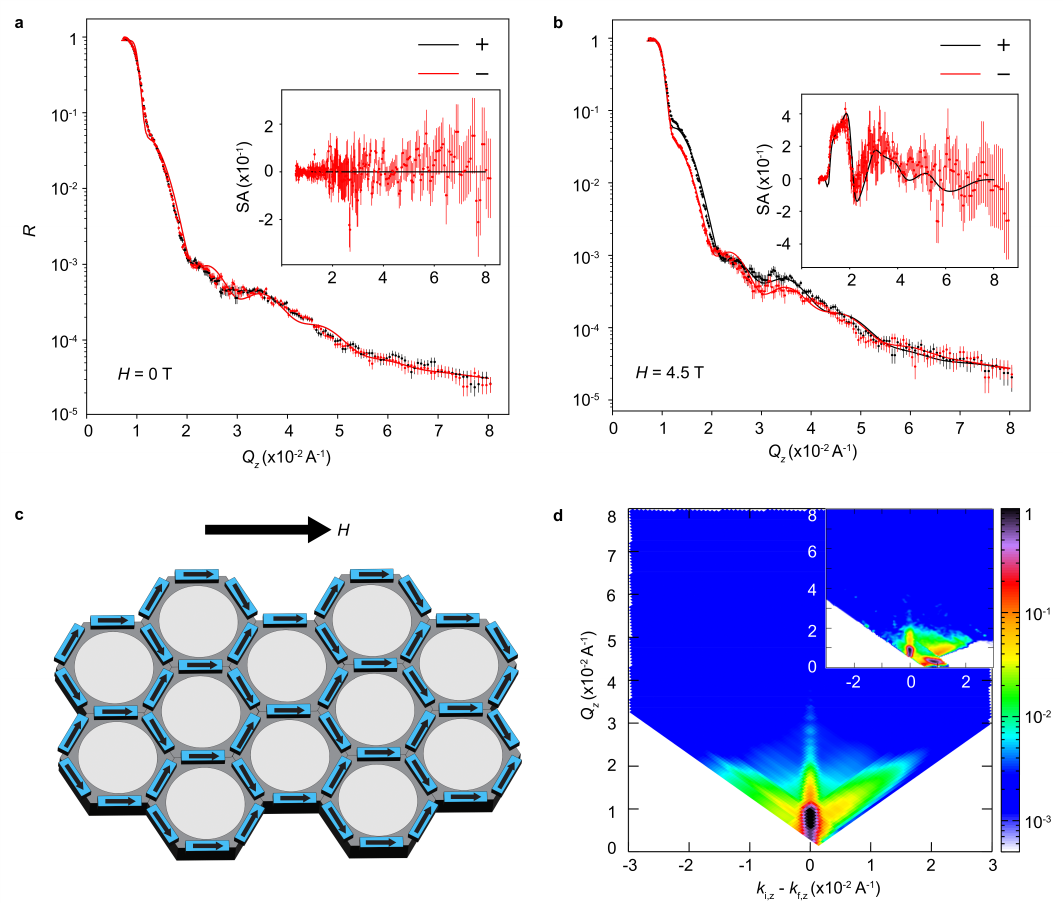}
\caption{\textbf{Magnetization study of Nd honeycomb in zero and finite applied field.} \textbf{a-b}, Spin polarized specular neutron reflectivity plots at $H=0$ and $H=4$ Tesla at $T = 4$ K. Overlapping reflectivity curves of opposite spin polarizations, $(+)$ and $(-)$, in zero field indicate the absence of net magnetization. Field application induces net magnetization by aligning Nd moments to the field direction, causing an irreversible split in the reflectivity curves. The inset shows the spin asymmetry (SA), indicating net magnetization. \textbf{c}, Magnetic moment configuration in applied field on the honeycomb motif. \textbf{d}, Numerically simulated off-specular reflectivity for the magnetic structure shown in {\bf c}. Inset: experimental data at $T = 5$ K. In the off-specular reflectometry graph, the $y$-axis represents the out-of-plane scattering vector $q_z= \frac{2\pi}{\lambda}\left[  \sin(a_i) + \sin(a_f) \right] $. The difference between the $z$-components of the incident and the outgoing wave vectors, $p_i - p_f= \frac{2\pi}{\lambda}\left[ \sin(a_i) - \sin(a_f)\right]$, is drawn along the $x$-axis, corresponding to the in-plane correlation. The simulation gives a good description of the experimental data. }
\label{fig3}
\end{figure*}

We further analyze the NSE results by plotting individual spin echo profiles, obtained at various neutron Fourier times and temperatures. As shown in \textbf{Figs. \ref{fig2}c-h}, the presence of a strong spin echo signal at all temperatures suggests paramagnetic scattering across the entire measurement temperature range, i.e., the Nd honeycomb lattice acts as a PM system. We have performed spin polarized neutron reflectometry (PNR) measurements on the Nd honeycomb lattice to verify this.\cite{Valeria} PNR measurements were performed on a 20$\times$20 mm$^2$ sample in a small guide field of $H$ = 20 Oe to maintain the polarization of incident and scattered neutrons (see Methods). In \textbf{Figs. \ref{fig3}a-b} we plot the specular intensity measured using spin-up $(+)$ and spin-down $(-)$ neutrons at $T = 5$ K in $H$ = 0 and 4.5 T fields. While there is no asymmetry between the $(+)$ and $(-)$ components in the specular data at $H$ = 0 T, it is prevalent in the applied field (see Figure S5 for $H = 1$ T data). The PNR data shows that the sample does not exhibit any net magnetization at $H$ = 0 T, but when subjected to a strong magnetic field the Nd atomic moments align along the field direction and develop a finite magnetization, as evidenced by the spin asymmetry analysis plot in the inset of \textbf{Fig. \ref{fig3}b}.
As shown in the spin polarized off-specular reflectivity plot in {\bf Fig. \ref{fig3}d}, a broad band of diffuse scattering along the $q_x$ direction is observed under the applied field, indicating an in-plane correlation between the field-induced magnetic moments in the honeycomb lattice.

The experimental data is modeled using the distorted wave Born approximation (DWBA) to infer the magnetic moment configuration (see Supplemental Information).\cite{BornAgain,Artur} As shown in \textbf{Fig. \ref{fig3}d}, the numerically simulated reflectometry pattern for magnetic moments aligned along the field direction is in good agreement with experimental data. The field induced moments impart $\pm Q$ charges to the honeycomb vertices. Similar measurements at $T = 300$ K in an applied field of $H = 4.5$ T field showed no traces of magnetism (see Fig. S6). Therefore, we can confirm that the Nd honeycomb lattice sample is PM at high temperature where thermal fluctuations overweigh the field-induced weak magnetization (below $T_N = 25$ K, Nd becomes AFM \cite{Bak}). In either case, the Nd honeycomb element cannot develop a net macroscopic moment along its length.

\begin{figure}\centering
\includegraphics[width=0.97\linewidth]{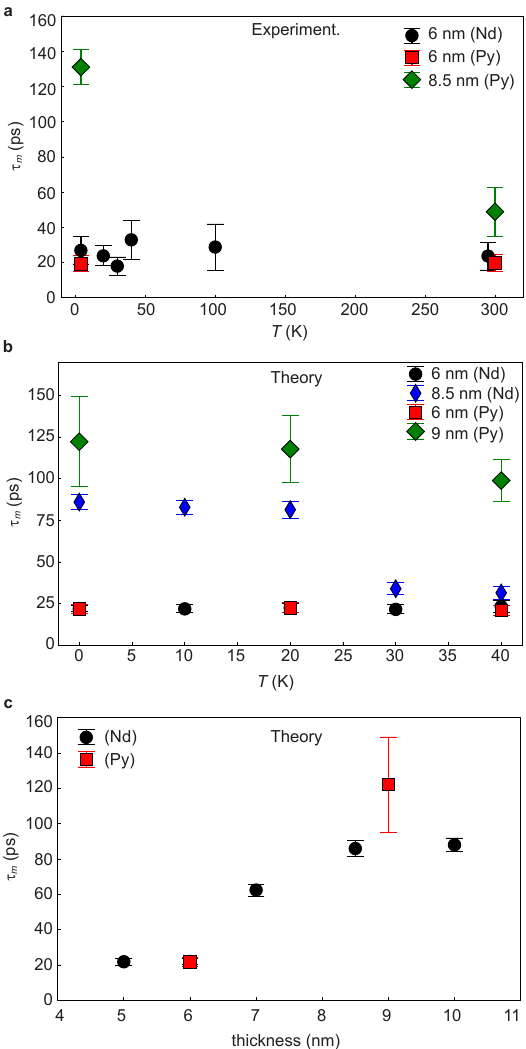}
\caption{\textbf{Self-propelled dynamic magnetic state in constricted honeycomb lattices due to quasi-particle kinetics.} \textbf{a}, Estimated experimental relaxation time $\taum$ versus temperature $T$ in Nd honeycomb lattice (black dots). For comparison, data from permalloy honeycombs of comparable (6 nm, red squares) and larger (8.5 nm, green diamonds) thickness is also shown. $\taum$ increases with thickness. Error bars are calculated from the square root of the variance of the least-square-fitted parameter. \textbf{b-c}, Theoretically estimated $\taum$ values as functions of thickness (b) and temperature (c). Theoretically obtained $\taum$ are in excellent agreement with experimental data. The data points represent the mean from multiple runs of stochastic simulation, and the error bars represent two standard deviations of the mean. The larger error bars on 9 nm Py are due to there being two distinct forms of motion observed in the model, with two quasiparticles moving in tandem traveling faster than isolated quasiparticles.}\label{fig4}
\end{figure}

Thus, the nature of the underlying magnetic state does not seem to influence the dynamic behavior. Next, to quantify the dynamic process we obtain the magnetic relaxation times $\taum$. For this purpose, we plot the normalized NSE spectral function, $S(q, t)/S(q, 0)$, versus Fourier time $t$ at several characteristic temperatures at $q = 0.06$ and 0.03 {\AA}$^{-1}$. The normalization is achieved by dividing the observed oscillation amplitude by the maximum measurable amplitude (see Supplemental Information), which is common practice in magnetic systems with unsettling fluctuations to the lowest measurement temperature (also see Fig. S4 and Supp. Information).\cite{Piotr} The normalized intensity reduces to the background level above the spin echo Fourier time of $\sim \! 0.5$ ns. This is the most general signature of a relaxation process in NSE measurements.\cite{Ehlers} An exponential fit of the NSE scattering intensity, $S(q, t)/S(q, 0)= C  \exp{(-t/\taum)}$, where $C$ is a constant, yields the relaxation time $\taum$ of the dynamic magnetic entity.\cite{Ehlers} The exponential fit gives a good description of the relaxation mechanism in Nd honeycomb lattices, see \textbf{Fig. \ref{fig2}{\bf f}-{\bf h}}. Even at the lowest temperature, the experimental data suggests a fast dynamic behavior, $\taum \sim 20$ ps. The estimated relaxation times of kinetic events at different temperatures are shown in \textbf{Fig. \ref{fig4}}. For comparison, we also show $\taum$ of magnetic charge defect relaxation in permalloy honeycomb lattices of varying thicknesses (6 nm and 8.5 nm) at low and high temperatures.

Several important observations can be immediately inferred from \textbf{Fig. 4a}. We notice that the dynamic behavior is prominent throughout the measurement temperature range in the Nd honeycomb lattice despite the absence of any external stimuli. The values of $\taum$ are comparable to those found in a 6 nm thick permalloy honeycomb lattice at $T = 300$ K as well as at 5 K: thus, the quantitative nature of the relaxation process is similar in both FM and PM/AFM nanostructured honeycombs. This is in direct contrast to the domain wall dynamics in nanostructured magnets, which requires magnetic field or current application. Additionally, the element size in permalloy honeycomb lattices is smaller than the domain wall size, $\sim 14$ nm.\cite{Coey,Huber} Therefore, the dynamic behavior in our nanoscopic lattice cannot be driven by domain wall kinetics. As the thickness of the nanomagnetic lattice increases to 8.5 nm, $\taum$ increases. The increment in $\taum$ is most significant at low temperature. This suggests that the dynamic behavior is thickness dependent. However, the relaxation time is still very fast and quantitatively comparable to or faster than the analogous behavior in bulk spin ice compounds, where it is attributed to the magnetic monopole kinetics.\cite{Ehlers,Collin}

To further elucidate this intriguing dynamic behavior, we now turn to theory. Nd has very interesting and rich electronic and magnetic properties: in the bulk form, it is AFM below $T_N$ with multiple ordering $k$ vectors, each with wavelengths on the order of several nanometers,\cite{Forgan} which is believed to be caused by Fermi surface nesting.\cite{Fleming}\cite{Dobrich} Nd thin films, on the other hand, exhibit a more complex glassy behavior below $T_N$, showing signs of long range disorder and nontrivial multiscale dynamics while retaining finite domains with local AFM order. At low temperatures the AFM order within these locally ordered domains is typically chosen from a discrete set of preferred ordering “Q-pockets” forming hexagons in $k$-space. The in-plane AFM ordering wavelength varies from 0.9 to 4.5 nm, with a dominant peak of roughly 3.7 nm.\cite{Kamber} Above $T_N$, it exhibits PM behavior.\cite{George}

In order to investigate the effect of imposing an artificial nanoscale honeycomb structure on the local AFM order found in Nd thin films, we undertook Landau-Lifshitz magnetization modeling. The “Q-pockets” are modeled via a $k$-dependent spin coupling with a hexagon of broad Gaussian peaks at the preferred wavelength of 3.7 nm. Specifically, $H = -\sum_k J_k \vec{s}_k \cdot \vec{s}_{-k}$ where $\vec{s}_k$ is the Fourier transform of the spin lattice $\vec{S}_i$, $J_k = J \sum_n e^{-(\vec{k}-\vec{q}_n)^2 /2 \sigma_q}$, with $\vec{q}_n = [\frac{2\pi}{\lambda}\cos (n\pi/3),\frac{2\pi}{\lambda}\sin (n\pi/3), 2\pi/c ] $, where $\lambda$ is the preferred in-plane magnetic ordering length set to 3.7 nm and $c$ is the out of plane unit cell length. This model results in a spin coupling that oscillates with distance, similar to the RKKY interaction, as one would expect from the nontrivial Fermi surface nesting found in Nd. Previously, the numerical study estimated an exchange coupling between nearby spins of the order of hundreds of $\mu$eV, which decays by several orders of magnitude at few lattice constants;\cite{Kamber} thus, exhibiting a gaussian-type decay, also found in our model. The magnetic dipole-dipole interaction between nearby Nd magnetic moments ($\sim 2.5$ Bohr magnetons) is of the order of tens of $\mu$eV, but decays more slowly, as $r^{-3}$. Hence, the dipole-dipole interaction is relevant to the dynamics of Nd nanomagnets. As such, we include dipole-dipole interactions between the discretized magnetic lattice points of the simulation. Thermal fluctuations are modeled with temperature dependent stochastic time steps as described in Ref. \cite{Leliaert}. 

For Py we use a similar model, but with joint anisotropy, a purely ferromagnetic spin exchange coupling, and a larger discretization length due to the larger size of quasiparticles. The strong thickness dependence for the Py case is found to be caused by a crossover in effective spin coupling strength relative to the anisotropy strength. For the thinner case (6nm) anisotropy dominates, whereas for the thicker case (9nm) the spin exchange coupling dominates (see Supp. Mat. for further information).

In our Nd nanostructure model, we find that the topological magnetic quasiparticles arise without the need of spin-orbit coupling (SOC) or Dzyaloshinskii-Moriya interactions (DMI). These quasiparticles form spiral textures around a small magnetic core, which points out-of-plane of the sample, with width of the order of $\lambda$. The structure of the quasiparticles is somewhat similar to half-skyrmions, as reported before,\cite{Jani} but the origin is quite different since here the vortex loop does not require SOC or DMI to stabilize it. The quasiparticles are topological in the sense that they have a topological index of $\frac{1}{4\pi} \int \vec{m} \cdot \left( \frac{d \vec{m}}{dx} \times \frac{d \vec{m}}{dy} \right) dxdy = \frac{1}{2}$, i.e. half sphere coverage. The out-of-plane half sphere coverage of these topological quasiparticles provides a way of releasing tension in frustrated magnetic configurations arising due to the nontrivial constricted geometry of the material.

In the case of bulk thin film geometry, these quasiparticles are rare and mostly appear on the boundaries between ordered domains with different AFM $\vec{k}$ orders (see Fig. S7 in Supp. Mat.). However, in the honeycomb structure with nanoscale element, the quasiparticles are much more common and appear to be stabilized by the restricting nature of the nanomagnetic geometry. The constricted nanostructure prevents ordered AFM domains from growing. The quasiparticles in the model relax at very fast time intervals. A similar behavior is detected in the constricted permalloy FM nanomagnetic lattice. However, the size of the topological defects is larger in the permalloy honeycomb lattice. {\bf Figs. 4b-c} show the plots of calculated quasiparticle relaxation time as functions of temperature and honeycomb lattice sample thickness. The simulation reproduces the key features of the temperature independent relaxation in 6 nm thick samples of Nd and Py honeycombs as well as the thickness dependence of relaxation times.

From the experimental and theoretical results shown here, we come to the conclusion that the dynamic behavior in constricted honeycomb lattices is mediated by vortex-type micro-spin profile quasiparticles with a net magnetic core, canting out perpendicular to the plane. Although a vortex spin profile has zero net magnetization, the core has finite magnetization and is hence detectable by the neutron probe. The quasiparticle floats across the lattice element with nearly barrier-less energy. Thus, the localized excitation in {\bf Fig. \ref{fig2}a-b} can be attributed to the kinetics of this quasiparticle. The size of the quasiparticle varies with the constituting material. As shown in {\bf Fig. 1e, f}, the vortex-type micro-spin configuration of the quasiparticle is limited to the exchange length or, the size of the nanomagnet component, whichever is smaller. The small size of the quasiparticle in Nd is also justifiable given the fact that Nd honeycomb lattice lacks net magnetization.

Our findings are expected to have general implications. The dynamic phenomena due to the magnetic quasiparticle kinetics should be detectable in any constricted nanomagnet where the width and thickness, or the diameter of a cylindrical shaped component, are smaller than the domain wall size of the magnetic material. Noting the lack of net moment formation along the connecting element in the Nd honeycomb, the length of the nanoscopic component is not relevant as the topological defect can be limited by the characteristic exchange length. Similarly, the nature of the underlying magnetic material is not important either. It will only affect the size of the topological quasiparticle defect and the relaxation rate. The finding can have significant implication to the technological advancement of nanoscopic devices.

\section{Methods}
\subsection{Sample fabrication}
The fabrication of the artificial honeycomb lattice involves the synthesis of a porous hexagonal template on top of a silicon substrate, deposition of Nd on top of the uniformly rotating substrate in a near-parallel configuration ($\sim$ 3$^{\circ}$) to achieve the 2D character of the system. The porous hexagonal template fabrication process utilizes diblock copolymer polystyrene (PS)-b-poly-4-vinyl pyridine (P4VP) of molecular weight 29K Dalton and volume fractions 70\% PS and 30\% P4VP, which can self-assemble into a hexagonal cylindrical structure of P4VP in the matrix of PS under the right conditions. A PS-P4VP copolymer solution of mass fraction 0.6\% in toluene was spin-coated on a polished silicon wafer (thickness 0.28 mm) at around 2300 rpm for 30 seconds, followed by solvent vapor annealing at $25^{\circ}$C for 12 hours. A mixture of toluene/THF (20/80 volume fraction) was used for the solvent vapor annealing. This process results in the self-assembly of P4VP cylinders in a hexagonal pattern in a PS matrix.  Submerging the sample in ethanol for 20 minutes releases the P4VP cylinders from the PS matrix, leaving a hexagonal porous template with an average hole center-to-center distance of 31 nm. This scaffolding is then utilized to create a magnetic honeycomb lattice by depositing Nd in a near-parallel configuration using E-beam physical vapor deposition, rotating the samples to achieve uniformity of the deposition. The resulting magnetic honeycomb lattice has a typical element size of about 11 nm length, 4 nm width, and controllable thickness. A typical atomic force micrograph of the honeycomb lattice is shown in Fig. 1a.

\subsection{Neutron spin echo measurements}
The Neutron Spin Echo (NSE) measurements were conducted on an ultrahigh resolution (2 neV) neutron spectrometer at beam line BL--15 of the Spallation Neutron Source at Oak Ridge National Laboratory. Time-of-flight experiments were performed using a neutron wavelength range of  3.5 -- 6.5 \AA~and neutron Fourier times between 0.06 and 1 ns. NSE is a quasi-elastic technique where the relaxation of a magnetic specimen is decoded by measuring the relative polarization change of the scattered neutron via the change in the phase current at a given Fourier time (related to neutron precession). To ensure that the detected signal is magnetic in origin, the experiment was carried out in a modified instrumental configuration of the NSE spectrometer where magnetism in the sample is used as a $\pi$ flipper to apply 180$^{\rm{o}}$ neutron spin flipping, instead of utilizing a flipper before the sample (see Fig. S2 in Supp. Information). Additional magnetic coils were installed to enable the xyz neutron polarization analysis for the calculation of $S(q, t)/S(q, 0)$.

NSE measurements were performed on a stack of 125 Nd honeycomb samples of $20 \times 20$ mm surface size and 6 nm thickness, to obtain good signal-to-background ratio. Considering the macroscopic separation distance between the honeycomb layers due to the 0.28 mm thick single crystal silicon substrate, inter-layer coupling between the stacked honeycomb samples is unlikely. The sample stack was loaded in a custom-made aluminum container inserted  in a close cycle refrigerator with 4 K base temperature. The sample stack was exposed to the neutron beam in the transmission geometry 3.9 m away from the detector and data was collected for 8 hrs on the average at each temperature and neutron Fourier time.

\subsection{Polarized neutron reflectivity measurement}
Polarized neutron reflectivity measurements were carried out on honeycomb samples of $20 \times 20$ mm surface size at the Magnetism Reflectometer beam line BL--4A of the Spallation Neutron Source at Oak Ridge National Laboratory. This instrument employs the Time-of-Flight technique in a horizontal scattering geometry with a bandwidth of 5.6~\AA~and a wavelength ranging from 2.6 to 8.2 \AA. Collimation of the neutron beam was conducted with slits and a 2D $^3$He area detector with 1.5 mm resolution was used at 2.5 m away from the samples. The samples were mounted on a copper cold finger of a closed cycle refrigerator with a base temperature of 5 K. Reflective supermirrors were used to polarize the beam, achieving better than 98\% polarization efficiency over the full wavelength band. A small guide field of $H$ = 20 Oe was used to maintain the beam polarization. The full vertical beam divergence was used for maximum intensity and a 5\% $\Delta\theta/\theta\sim\Delta q_{\rm z}/q_{\rm z}$ relative resolution was used in the horizontal direction. Data was collected at various temperatures and applied in-plane magnetic fields.

\subsection{Specular reflectivity fitting and off-specular scattering modeling}
The fitting of the specular reflectivity was carried out using the Distorted Wave Born Again (DWBA) approximation as implemented in the BornAgain software, version 20. The basic procedure outlined in Ref. \cite{BornAgain} was adopted, with global minimum searching via a differential evolution algorithm followed by least-square local minimum searching. An offset in $q_z$ was also fitted in order to correct for the systematic shift of $q_z$ from instrument misalignment. The reflectivity data collected at $T$ = 5 K and $H$ = 4.5 T was first fitted by varying both structural and magnetic parameters. The resulting structure data was then fixed during the fitting for the reflectivity data collected at $T$ = 5 K, $H$ = 0 T and 1 T while allowing magnetic parameters to vary.

For the modeling of the off-specular scattering intensity, a multilayer honeycomb structure model was constructed from the refined structural parameters obtained from the specular reflectivity fitting after correcting for the surface density. The model consist of a bottom layer of silicon substrate followed by the three structured layers of PS, Nd and oxide sequentially in honeycomb geometry. The top oxide layer was introduced to account for any oxidation effect that might cause a magnetically dead region with different nuclear scattering length density compared to pure Nd. For the honeycomb structured layers, we introduced a hexagonal lattice with lattice parameter 31 nm out of cylindrical cut-outs with a cylinder radius of 11.2 nm. A 2D lattice interference function was adopted to account for imperfection of the self-assembled system with a coherence length of 1000 nm. Magnetic moments in the Nd layer are represented as rectangular bars with 11 nm $\times$ 4 nm (length $\times$ width) on the honeycomb edges with their orientations as presented in Fig. 3c, considering the experiment condition ($H$ = 4.5 T) as well as the net magnetization revealed in specular reflectivity data fitting. The magnetization of the rectangular bar magnets was determined from Nd atomic moments according to Ref. \cite{Nd_atomic_moment_paper}.


\section{Acknowledgements}

We thank Antonio Faraone and Georg Ehlers for helpful discussion on the use of NSE technique in probing magnetic materials and designing the experiment. This work was supported by the U.S. Department of Energy, Office of Science, Basic Energy Sciences under Awards No. DE-SC0014461 (DKS) and  DE-SC0019109 (CAU). This work utilized the facilities supported by the Office of Basic Energy Sciences, US Department of Energy.

\section{Authors Contributions}

D.K.S and C.A.U. jointly led the research. D.K.S. envisaged the research idea and supervised every aspect of the experimental research. C.A.U supervised theoretical research. Samples were synthesized by J.G. PNR and NSE measurements were carried out by J.G., VL, L.S., P.Z, D.K.S.  Analysis was carried out by J.G., V.L., L.S., P.Z. and D.K.S. Theoretical calculations were carried out by D.H. and C.A.U. The paper was written by D.K.S. and C.A.U. with input of  all co-authors.

\section{Additional Information}

Supplementary Information is available for this paper.
Authors declare no competing financial interests. Correspondence and requests for materials should be addressed to D.K.S.

\end{document}


\baselineskip 18pt
\parskip0pt
\parindent8mm

\begin{center}
{\large Supplementary Information}\\
{\bf \Large Emergent topological quasiparticle kinetics in constricted nanomagnets}\\
J. Guo, D. Hill, V. Lauter, L. Stingaciu, P. Zolnierczuk, C. A. Ullrich, and D. K. Singh
\end{center}

\section{Neutron Spin Echo measurement and data analysis}
\subsection{Modified instrument for paramagnetic NSE}
\begin{figure}
    \includegraphics[width=\textwidth]{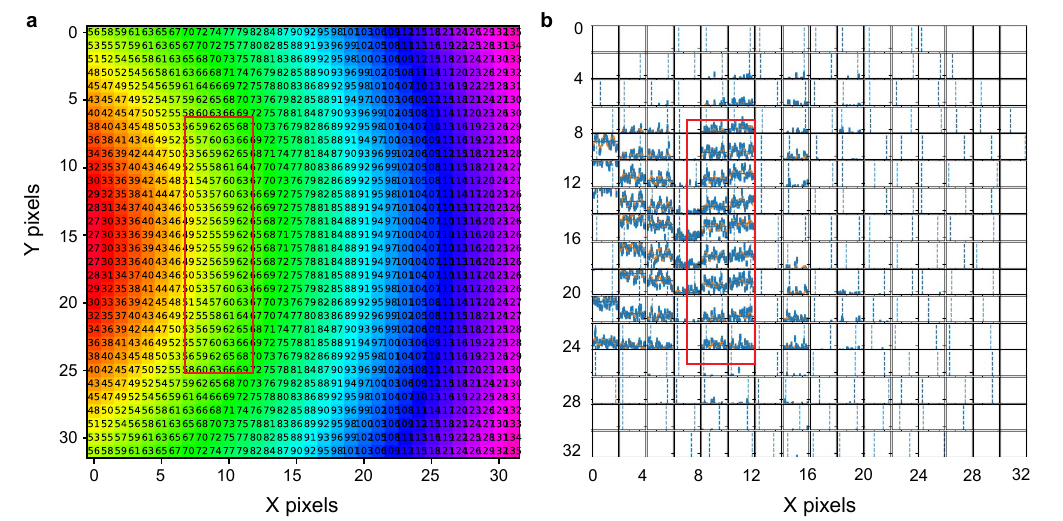}
    \caption{\textbf{The detector $\boldsymbol{q}$ distribution and echo signals of the Nd honeycomb lattice measured at $\boldsymbol{T}$ = 295 K.} \textbf{a} The flux-weighted $q$ distribution across the whole detector in ToF tbin 17, corresponding to a Fourier time of 0.02 ns, the numbers in the individual pixels denote the $q$ value in the unit of 0.001 \AA$^{-1}$. \textbf{b} The echo signals in ToF tbin 17 on the detector. Every 2 pixels along X and Y directions are grouped for the clearness of the demonstration. The red rectangles mark the area for X, Y pixel grouping for furthur data analysis with $q \sim$ 0.06 \AA$^{-1}$.} 
    \label{fig:suppl_q_map}
\end{figure}

The Neutron Spin Echo (NSE) measurements were conducted on an ultrahigh resolution (2 neV) neutron spectrometer with a dynamic range 1 ps to 300 ns at beam line BL--15 of the Spallation Neutron Source, Oak Ridge National Laboratory. A 3 {\AA} wavelength bandwidth of neutrons (3.5--6.5 \AA) was used in the experiment, and the scattered neutrons were collected with a 30 cm $\times$ 30 cm position-sensitive $^3$He detector 3.9 m away from the sample. The detector has 32 $\times$ 32 X, Y pixels laterally to keep track of the scattering angles of the detected neutrons. It also has 42 time of flight (ToF) channels (tbins) that label the time frame of the detected neutron which encodes the detected neutron's wavelength. Thus, the data collected by the detector is a 32 $\times$ 32 $\times$ 42 array. By carefully grouping the X, Y pixels and tbins, one can extract echo signals at different Fourier times $t$ [see Eq. (\ref{eq:fouriertime}) below for the definition] and $q$ values from a single echo measurement. Note that for a 2D area detector with ToF, each detector pixel corresponds to a broad range of $q$ arising from the wavelength band of neutrons from the source. In this sense, the 42 ToF tbins essentially define the resolution of the wavelength of the detected neutrons. Suffice to say that at a given X, Y detector pixel, within a single tbin, the collected neutrons are assumed to have the same wavelength, hence the same $q$ and Fourier time $t$. The scattering vector $q$ also varies a cross to the detector area due to the change of scattering angle. For illustration purposes, in Fig.~\ref{fig:suppl_q_map}, we have presented the $q$-assignment of the detector pixels within a single tbin corresponding to a Fourier time $t$ = 0.02 ns.

Paramagnetic NSE requires separate measurements for the echo signal and the total magnetic scattering for the normalization purpose. In the echo signal measurement, the scattered beam intensity is measured as one scan through the phase current that introduces the field integral asymmetry, and no field is applied around the sample. Instead of using a $\pi$ flipper before or after the sample to flip the neutron's polarization as is typically done for soft matter application of NSE, here the magnetic sample itself does a $\pi$ flip of the neutron's polarization due to the nature of the interactions between magnetic moments and neutron's polarization. Thus, the $\pi$ flipper was removed in the echo signal measurement. This way, we are certain that any scattering effects that cannot invoke a $\pi$ flip to the neutrons' polarization (including structural scattering) will not contribute to an echo signal but only a flat background~\cite{pappas2006}.  Echo signals at multiple temperatures and nominal Fourier times (the Fourier time associated with the neutron of the maximum wavelength) were measured.

\begin{figure}
\centering
    \includegraphics[width=0.9\textwidth]{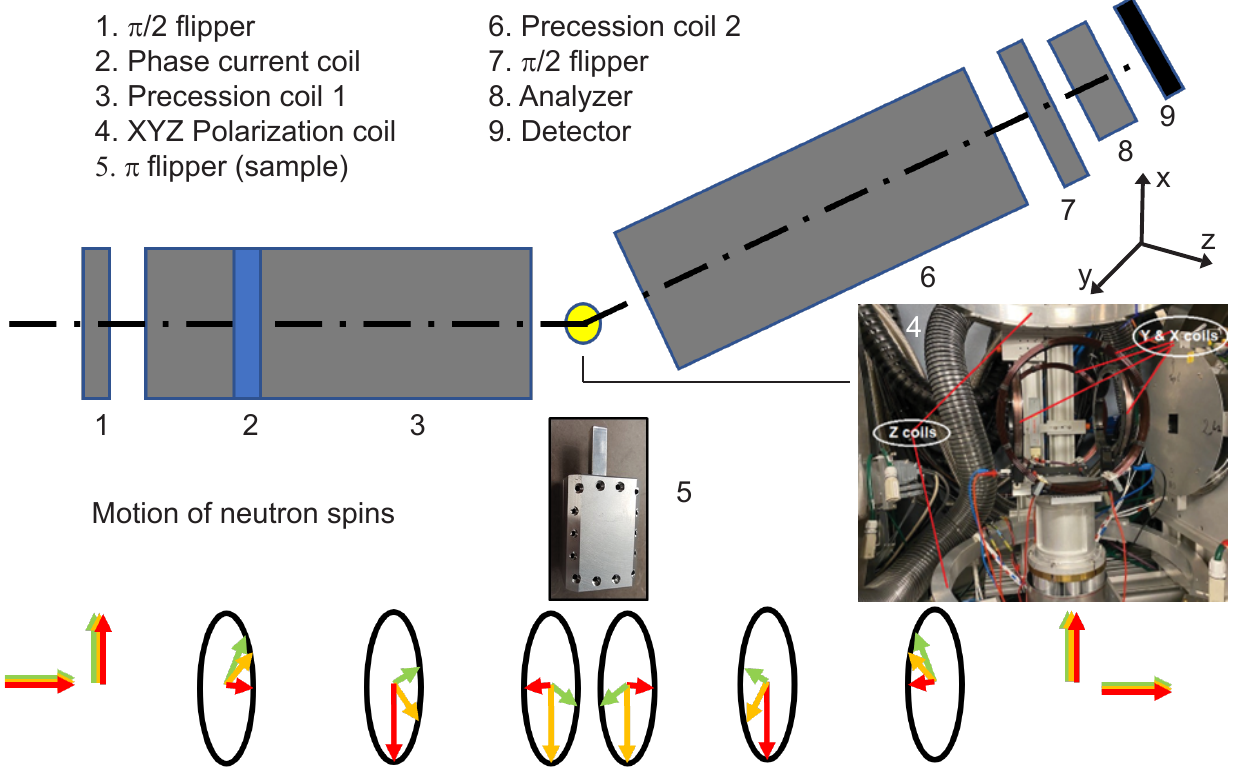}
    \caption{\textbf{Design of modified NSE instrument, utilized for the paramagnetic NSE experiment.} Inset shows the stack of samples sealed in an aluminum can, which serves as a $\pi$ flipper. Additional magnetic coils are installed around the sample to enable the polarization analysis. }
    \label{fig:supp_instrument}
\end{figure}

The total magnetic scattering is measured in the 3D xyz polarization analysis where a small guide field is applied around the sample. For this purpose, three sets of Helmholtz-style coils were orthogonally mounted around the sample position for $x$, $y$ and $z$ orientations, see inset in Fig.~\ref{fig:supp_instrument} for details of the coil installment. Six additional polarization measurements along $x$, $y$ and $z$ directions ($x_{\mathrm{up}}, x_{\mathrm{dn}}, y_{\mathrm{up}}, y_{\mathrm{dn}}, z_{\mathrm{up}}, z_{\mathrm{dn}}$) were performed at each temperature with a small guide field $\sim$ 10 Oe applied around the sample to define the quantization direction for the neutron spin and maintain its polarization. A magnetic needle was used to make sure the polarization field direction aligns to the $x$, $y$ and $z$ orientations. Here, $z_{\mathrm{up}}$ and $z_{\mathrm{dn}}$ for instance, were measured as the spin-non-flip scattering (SNF) and the spin-flip scattering (SF) cross sections with the field along the positive $z$ direction. The polarization analysis was performed according to the method described by Pappas et al.~\cite{pappas2006} with the currents in the coils tuned following Ehlers et al. \cite{ehlers2013}.

To enhance the signal-to-background ratio, a stack of 127 samples of about 20 $\times$ 20 mm$^2$ was loaded in a custom-made aluminum sample container. The Si substrate for each sample is $\sim$ 0.28 mm in thickness. Thus, no inter-layer couplings between the magnetic honeycomb layers need to be considered. The sample container was then inserted into a close cycle refrigerator with a base temperature of 4 K with the sample stack exposed to the neutron beam in the transmission geometry such that the neutron beam direction is parallel to the sample normal direction. A schematic diagram of the modified NSE instrument is presented in Fig. \ref{fig:supp_instrument}.

\subsection{Analysis of the Neutron Spin Echo data}
NSE experiments probe the intermediate scattering function of the sample studied, given by~\cite{Mezei2003}
\begin{equation}
    S(q,t)=\int{\cos(\omega t)S(q,\omega)d\omega},
\end{equation}
where $S(Q,\omega)$ is the scattering function of the sample and $t$ is the Fourier time, defined as
\begin{equation}
    t=J\lambda^3\frac{\gamma_{\mathrm{n}}m_{\mathrm{n}}^2}{2\pi h^2}, \label{eq:fouriertime}
\end{equation}
with $\gamma_{\mathrm{n}}$, $m_{\mathrm{n}}$ and $\lambda$ being the neutron's gyromagnetic ratio, mass and wavelength, and $h$ denotes Planck's constant. $J=\int{|B|dl}$ is the magnetic field integral along the neutron's path through the precession coil \cite{piotr2019}. The field integrals are designed to be the same in both precession coils before and after the sample such that the precession phase acquired by the neutron in the first coil will be exactly recovered at the end of the second coil, given that the neutron does not change its velocity while interacting with the sample. For quasi-elastic scattering, the neutron gains a net phase at the end of the second coil before it reaches the analyzer. By systematically stepping through an additional phase current (convert to an additional field integral) in either the first or the second coil, a cosine modulation of the sample's intermediate scattering function is realized, which is the typical raw data, echo signal, one would analyze in an NSE experiment.

\subsubsection{Raw echo signal analysis}
The echo signal intensity for a given neutron wavelength $\lambda$ has a cosine function shape given by
\begin{equation}
    I(\phi) = A\cos(\phi \lambda) + B,
\end{equation}
where $\phi=dJ \gamma_{\mathrm{n}}m_{\mathrm{n}}/h$ and $dJ$ is the phase asymmetry between the two precession coils introduced via the scanning phase current. When signals from neutrons of a certain wavelength span $\lambda_{\mathrm{avg}}\pm d\lambda$ are added up, the total intensity resembles a cosine function modulated by an envelope function
\begin{equation}
    I(\phi)=A\cos(\phi \lambda_{\mathrm{avg}})\frac{\sin(\phi d\lambda)}{\phi d\lambda}+B. \label{eq:echo}
\end{equation}

\begin{figure}
\centering
    \includegraphics[width=\textwidth]{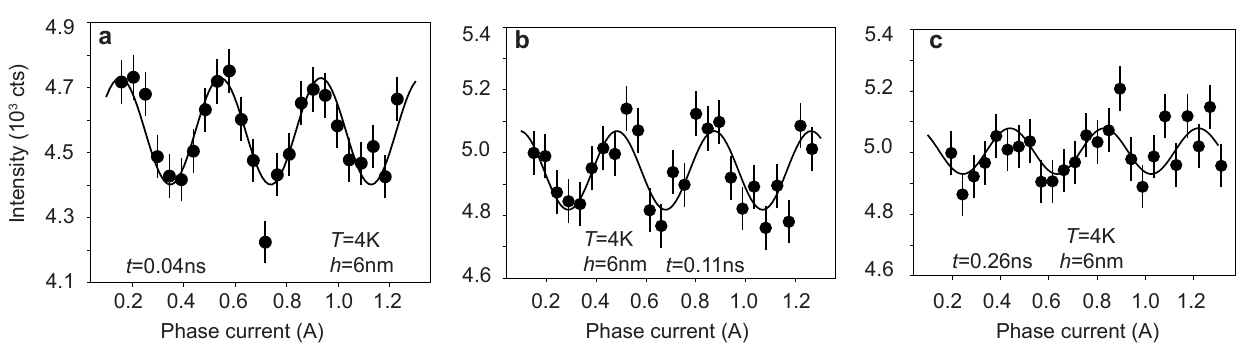}
    \caption{\textbf{Echo profiles of the Nd honeycomb lattice at different Fourier times measured at $\boldsymbol{T}$ = 4K.} \textbf{a} $t$ = 0.04 ns. \textbf{b} $t$ = 0.11 ns. \textbf{c} $t$ = 0.26 ns. The Fourier time of the echoes are labeled on the graphs. Clearly, as the Fourier time increases, the echoes exhibit smaller amplitudes with worse sinusoidal profiles. In all plots, the error bar represents one standard deviation.}
    \label{fig:supp_4K_echoes}
\end{figure}

The echo intensities were obtained by adding up signals in X, Y detector pixels and then summing over different ToF tbins. Grouping of the X, Y detector pixels were guided by the false color detector image as well as the detector pixel image (Fig. \ref{fig:suppl_q_map}) to center around the intense scattering signal. The echo profiles presented in Fig. 2 were extracted from a single ToF tbin by summing over detector area of 5 $\times$ 18 in X and Y directions (X = 7 to 11, Y = 7 to 24) with the mean neutron wavelength of 4.6~\AA~in an effective wavelength band of 0.07~\AA. We present in Fig.~\ref{fig:supp_4K_echoes} echo profiles measured at $T$ = 4 K with increasing Fourier time $t$. It is clear that as the Fourier time increases, the echoes exhibit smaller amplitudes with worse sinusoidal profiles.

\subsubsection{Calculation of the intermediate scattering function}

The intermediate scattering function $S(q,t)/S(q,0)$ as a function of the Fourier time $t$ (Fig. 2) can be calculated from the fitted echo amplitude. For non-magnetic samples,
\begin{equation}
\frac{S(q,t)}{S(q,0)}=\frac{2A}{U-D} \label{eq:sqt_UD}
\end{equation}
where $A$ denotes the fitted amplitude of the echo signal, $U$ and $D$ denote the spin up (non-$\pi$ flipping) and spin down ($\pi$ flipping) measurement of direct scattering without precession. $U-D$ measures the maximum obtainable echo amplitude and is used as the normalization factor. In paramagnetic NSE with a magnetic sample, half of the magnetic scattering intensity $M/2$ is used for normalization~\cite{pappas2006}. $M$ and $U, D$ are calculated as
\begin{equation}
    M=2(z_{\mathrm{up}}-z_{\mathrm{dn}})-[(x_{\mathrm{up}}-x_{\mathrm{dn}})+(y_{\mathrm{up}}-y_{\mathrm{dn}})],
\end{equation}
\begin{equation}
        U=\frac{x_{\mathrm{up}}+y_{\mathrm{up}}+z_{\mathrm{up}}}{3}, D=\frac{x_{\mathrm{dn}}+y_{\mathrm{dn}}+z_{\mathrm{dn}}}{3}.
\end{equation}
The intermediate scattering function is then determined as
\begin{equation}
    \frac{S(q,t)}{S(q,0)}=\frac{4A}{M}. \label{eq:sqt}
\end{equation}
In reality, the weak magnetic scattering signal amid the intense structure scattering signal prohibits a statistically reliable determination of $M$. Eq.~(\ref{eq:sqt}) thus leads to large fluctuation in $S(q,t)/S(q,0)$. Therefore, in this work, we have calculated the intermediate scattering function using Eq.~(\ref{eq:sqt_UD}) and corrected with the average ratio of  $2(U-D)/M$ of the three ToF tbin groups at each temperature. The same X, Y pixels (as used for echo profile demonstration) and ToF tbins selection was used for the determination of both the echo amplitude and the normalization factor. At each temperature, independent echo measurements were performed at 5 nominal Fourier times (0.06 ns, 0.1 ns, 0.3 ns, 0.7 ns, 1.0 ns) by setting different values for the field integral $J$. From each echo measurement at a given nominal Fourier time, 3 echo signals of different Fourier times were extracted by grouping ToF tbins, T = 10 to 19, T = 20 to 29, T = 30 to 39, respectively, with the phase shift along the tbins corrected. These echo signals were normalized to the neutron flux and were further analyzed for the intermediate scattering function calculation.

\begin{figure}
\centering
    \includegraphics[width=0.9\textwidth]{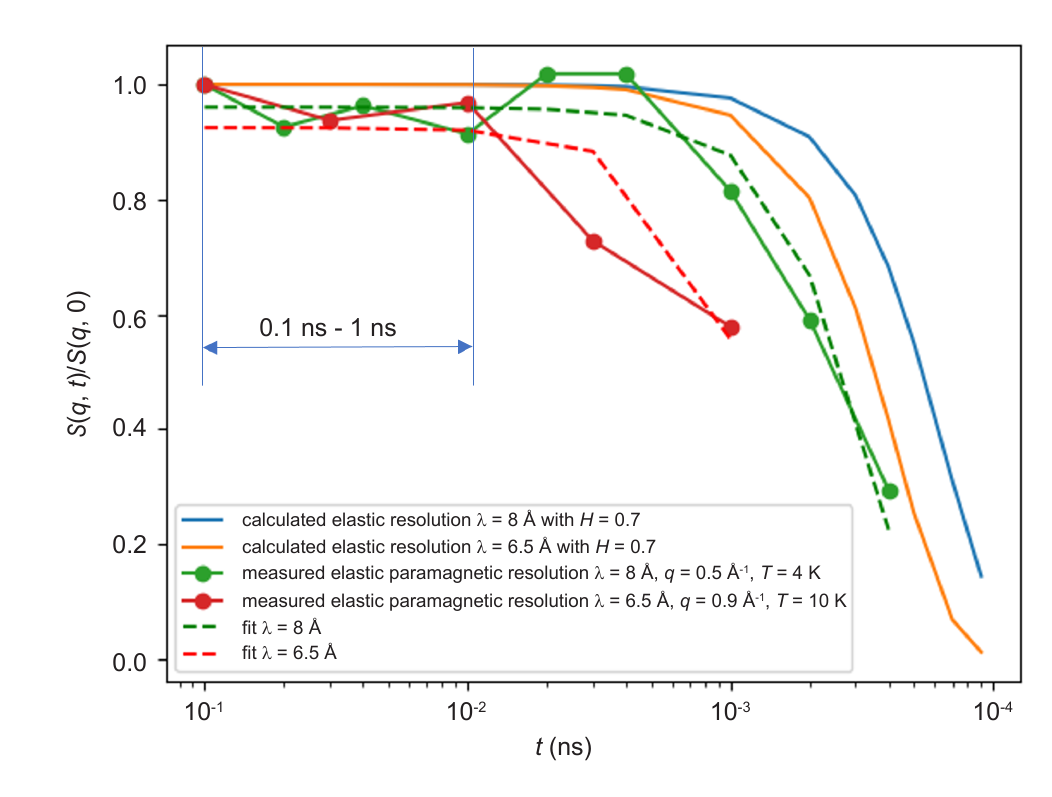}\vspace{-4mm}
    \caption{\textbf{Measured paramagnetic resolution function at SNS-NSE and simulated elastic resolution function in soft matter regime.} The paramagnetic resolution functions are measured at different $q$ values on a Ho$_2$Ti$_2$O$_7$ sample at temperatures below its frozen temperature $T$ = 20 K, with two different neutron wavelengths. $H$ represents the relative field integral homogeneity of the SNS-NSE spectrometer, a calculated value specific for SNS-NSE. Clearly, the paramagnetic resolution function below Fourier time $t <$ 1 ns is predominantly flat. Thus, any resolution correction to the measured intermediate scattering function in this work will not affect the relaxation time extracted from the exponential fitting, but only shift the data on the y-axis.}
    \label{fig:NSE_resolution}
\end{figure}

Unlike experiments conducted with other spectrometers, the measured dynamic structure factor $S_{\mathrm{exp}}(q,\omega)$ is a convolution between the true dynamic structure factor
$S(q,\omega)$ and the instrument resolution $R(q,\omega)$, that is $S_{\mathrm{exp}}(q,\omega)=S(q,\omega)\ast R(q, \omega)$. NSE directly probes the intermediate scattering function, which is the Fourier transform of the dynamics scattering function  $S(q,t)=\int{\cos(\omega t)S(q,\omega)d\omega}$, thus $S_{\mathrm{exp}} (q,t)=S(q,t)R(q,t)$. Therefore, in an NSE experiment, the resolution function can be simply divided out to obtain the true intermediate scattering function. At SNS-NSE the resolution of the instrument and elastic scattering contribution is assessed by measuring a perfect elastic scattering sample, usually mounted in the same container as the sample to investigate, measured over the same $q$ range, Fourier time range, and wavelength. For soft matter measurements, solid graphite and Al$_2$O$_3$ as well as TiZr are usually used depending on the scattering angles. For paramagnetic NSE measurements, Ho$_2$Ti$_2$O$_7$, a well-known classical spin ice material frozen below $T$ = 20 K is used where it exhibits no dynamics. However, for the measurement performed in this work, it was not possible to use Ho$_2$Ti$_2$O$_7$ sample as the resolution since it only scatters and produces reliable echoes at high scattering angles, while the magnetic honeycomb samples scatter in the small-angle regime. Another common practice in quasi-elastic techniques is to measure the same sample at $T \le$ 4 K where presumably all dynamics freezes. However, it is well observable in our measurements that even at $T$ = 4 K fast dynamics still exists. Therefore, we decided not to reduce the data by elastic resolution, and the relaxation times $\uptau_{\mathrm{m}}$ were extracted directly from fitting the intermediate scattering function at each temperature. In support of our decision is the fact that at SNS-NSE below Fourier times $t <$ 1 ns, the elastic resolution is predominantly flat (linear). This means that any reduction by resolution will not affect the relaxation observed in the data but only the intensity scaling on y-axis. Fig.~\ref{fig:NSE_resolution} demonstrates two measured Ho$_2$Ti$_2$O$_7$ resolutions from previous paramagnetic measurements in the beam line at $T$ = 4 K and 10 K, for 2 different wavelength 6.5 \AA~and 8 \AA. These experimental elastic resolution magnetic data are presented, together with their fits and simulated resolution in the soft-matter regime. One can easily observe the linear behavior in the range of $t <$ 1 ns, which is the predominant range of our measurements. In this sense, the conclusion of a thermally independent relaxation process that persists in the Nd honeycomb sample will still be valid without instrument resolution correction since the correction equally applies to each temperature. Ultimately, it is the estimation of the relaxation time from $S(q,t)/S(q,0)$ that is of the most scientific importance.

\section{Theory}
\subsection{Nd Model}
The Hamiltonian used to model the magnetic dynamics of the Nd nanostructure is
\begin{equation}
    H = \sum_{\vec{k}} J_{\vec{k}}  \vec{s}_{\vec{k}} \cdot \vec{s}_{-\vec{k}} + H_{d-d} \:,
\label{eqn:ham}
\end{equation}
where $\vec{s}_k$ is the Fourier transform of the magnetization lattice $\vec{M}_i$. The coupling is a sum over Gaussian peaks, $J_{\vec k} = J \sum_n e^{-(\vec{k}-\vec{q}_n)^2 /2 \sigma_q^2}$, with preferred AFM ordering vectors $\vec{q}_n = [\frac{2\pi}{\lambda}\cos (n\pi/3),\frac{2\pi}{\lambda}\sin (n\pi/3), 2\pi/c ] $, where $\lambda$ is the preferred in-plane magnetic ordering length set to 3.7nm and $c$ is the out of plane unit cell length, which is included to account for the antiferromagnetic coupling between atomic layers. $\sigma_q$ is a measure of the width of the energy minima in $k$-space. $H_{d-d} = \sum\frac{\mu_0}{4\pi}[\vec{M}_i \cdot \vec{M}_j -3 (\vec{M}_i \cdot \hat{r}) (\vec{M}_j\cdot \hat{r})]/r^3$ denotes the magnetic dipole-dipole interaction, which, as discussed in the main text, is likely to be relevant on the nanometer length scale for Nd. Replacing the dipole-dipole interaction term with a shape anisoptropy approximation results in a static AFM state with no quasiparticles, so we emphasize that $H_{d-d}$ is essential for capturing the itinerant physics observed in the experiment.

The first term in the Hamiltonian (\ref{eqn:ham}) applies the methods of ref. \cite{Yambe2022} to phenomenologically capture the six-fold symmetric ordering preference observed in elemental Nd \cite{Forgan1989} \cite{McEwen1985}, particularly in the thin film case \cite{Kamber2020,Verlhac2022}. The broadening of the Hamiltonian minima, represented by the finite $\sigma_q$, is included to account for the broadening of the ordering ``Q-pockets'' and the short range of the magnetic exchange coupling, both observed in Ref. \cite{Kamber2020}.

\subsection{Py Model}

In order to model Py, we employ ferromagnetic coupling between joints and vertices, with one spin per each, respectively, as well as a local easy-axis anisotropy on each joint. In Py the quasiparticles are much larger, with the out of plain spin texture region covering a whole vertex, which allows for a larger discretization.

The Hamiltonian is
\begin{equation}
    H= -\sum_{\langle i,j \rangle} J \vec{M}_i \cdot \vec{M}_j - \sum_{\text{joints}} K \left[\vec{M}_i \cdot \hat{e}(\vec{r}_i)\right]^2.
\end{equation}
We must have $J,K>0$ and $K>J$ in order to get the expected spin ice behavior with a preference for 2-in-1-out and 1-in-2-out configurations, and quasiparticle excitations in the form of 3-in or 3-out configurations.

We use a larger geometry which includes several hexagons of the honeycomb structure in the Py model, as shown in Fig. \ref{fig:theory}e. This choice deviates from the geometry of the two-vertex model used for Nd, because employing a similar geometry for the Py model failed to capture the significant increase in $\tau$ in going from 6nm to 8.5nm. In order to account for thickness variation, the honeycomb lattice is nearest-neighbor coupled to vertically shifted copies of itself. Specifically, we employ two layers and three layers to represent thicknesses of 6nm and 9nm, respectively.

Merely increasing the thickness by itself is insufficient to model the large $\tau$ difference observed in the experiment. However, we can induce a large jump in $\tau$ in the model by setting $K$ sufficiently small that increasing thickness causes the \textit{effective} joint-vertex coupling strength to surpass the anisotropy strength. By this we mean that the thicker multi-layer system begins to behave akin to a single-layer system with $K<J$. This was found to be the case for parameters in the vicinity of $K \sim 2J$. For models satisfying $K \sim 2J$, the two layer system behaves as we expect, but in the three layer system the spins on joints are allowed to cant significantly away from their preferred axis, see Fig. \ref{fig:theory}f. This canting results in longer range correlations between spins. As a result, moving a 3-in or 3-out quasiparticle causes movement of spins multiple vertices away: in this case, the quasiparticle is effectively coupled to more spins. Consequently, displacing a 3-in or 3-out quasiparticle results in movement of spins much further away than in the typical spin ice case. In this scenario, the quasiparticle becomes effectively coupled to more spins, which causes a drag on the quasiparticle's motion and thus a spike in $\tau$. 

The quasiparticles are able to counter this drag effect to some degree by pairing and moving in tandem. This pairing effect provides a distinct form of motion which leads to a double peaked distribution of speeds, which is why the error bounds in Fig. 4b,c on the 9 nm Py points are much broader than the other error bounds.

Even with fixed parameters at $K = 2J$ and varying thickness, the model still falls just short of fully replicating the significant $\tau$ difference observed experimentally. However a model with a reduced anisotropy, $K=J$, only in the three layer case was able to reproduce the large spike, and this is the model we used in our final analysis.

A degree of reduced shape anisotropy is anticipated due to the changing shape of the joints resulting from increasing the thickness of the artificial honeycomb. Quantitatively predicting the extent of change in anisotropy energy due to shape alteration is difficult. Nonetheless, studies such as Ref. \cite{Pastor} have demonstrated that transitioning from a one-dimensional to a two-dimensional metallic ferromagnetic system can reduce anisotropy by an order of magnitude, and that the anisotropy energy can exhibit significant oscillations as the shape varies in intermediate regimes. Both shape anisotropy and magnetocrystalline anisotropy can vary substantially as a function of shape in nanostructures \cite{Munoz}. Shape anisotropy changes most rapidly in the range of aspect ratios from 1 to 2 \cite{Bance}, and our Py systems change in aspect ratio (between joint axis and out of plane axis) from 1.83 (11nm/6nm) to  1.29 (11nm/8.5nm). Therefore, a 2-fold change in anisotropy is well within reason for the Py systems under consideration here.

\subsection{Numerical simulation details}

The time propagation is modeled by the Landau-Lifshitz equation
\begin{equation}
\frac{d\vec{M}}{dt} = - \gamma \left( \vec{M} \times \vec{B}_{\text{eff}} - \alpha \vec{M} \times (\vec{M} \times \vec{B}_{\text{eff}}) \right),
\end{equation}
where $ \vec{B}_{\text{eff}}(\vec{r},t) = \frac{\delta H}{\delta \vec{M}}$ is the effective magnetic field, $ \gamma $ is the gyromagnetic ratio, and $\alpha $ is the unitless damping parameter. The damping parameter was set to $\alpha =0.1$ for the main time propagation calculations. A higher value of $\alpha$ was used when searching for an initial state, which was found by heavily damping a random starting configuration. The magnetization is discretized $\vec{M}(\vec{r},t) \rightarrow \vec{M}(\vec{r_i},t_n)$. The grid size used for the Nd model was $\Delta x = 0.5$nm and the time step size was $\Delta t = 0.36$ps. Time steps are evaluated with the 4th order Runge-Kutta method. The value of $\Delta x$ was chosen in order to be sufficiently small to resolve the AFM order and topological quasiparticles in the spin texture. $\Delta t$ was set by checking for numerical stability of the normalization of $\vec{m}_i$. The geometry used for Nd calculations is depicted in Fig. 1f of the main text, with the in plane dimensions fixed and out of plane size varied for different widths, and free boundary conditions were used. The geometry used for modeling Py is shown in Fig. \ref{fig:theory}e,f. Both the Nd and Py models used on the order of 1000 sites per layer.

For finite temperature modeling, thermal fluctuations $\vec{B}_{\text{th}}$ are modeled with a Gaussian distribution with standard deviation given by \cite{Leliaert2017}
\begin{equation}
    \sigma_{\text{th}} = \sqrt{\frac{2k_B T \alpha}{M_s \gamma \Delta x^3 \Delta t}} \:.
\end{equation}
This expression leads to a Gaussian distributed torque on the magnetization. For the Nd model, the change in normalized magnetization ($\vec{m} = \vec{M}/M_s$) per time step has a standard deviation of approximately 
\begin{equation}
\delta m \sim \sqrt{\frac{T}{320\text{K}}} \:.
\end{equation}
This expression suggests a breakdown of the model near room temperature, as the thermal fluctuations for a given time step reach the same order of magnitude as the magnetization itself. The regime of applicability could potentially be pushed higher by using a smaller timestep, as the thermal change in magnetization scales as $\delta \vec{m} \sim \sigma_{\text{th}} \Delta t  \sim O(\sqrt{\Delta t})$, however this model is likely to become fairly inaccurate somewhere near the Néel temperature of Nd in any case, so, instead of adopting a smaller time step, we simply restrict ourselves to simulating $T \ll 100$K with this model. There was no such limitation on the Py model due to the much larger discretization cell volume, but no interesting features were found at higher temperatures, and accurate measurements of $\tau$ were much more difficult due to increased noise, so this data was omitted.

By fitting the Nd model to the experimentally observed time scale on the order of 20ps and assuming a magnetic moment per lattice site of $\mu \sim 2.5 \mu_B$, we estimate the spin-spin nearest neighbor coupling energy to be on the order of a few hundred $\mu$eV, in line with previous studies, while the magnetic dipole-dipole coupling is on the order of ten $\mu$eV, but the latter has a much longer range, which is why we consider it necessary to include dipole-dipole interactions in the model.

\clearpage

\clearpage

\clearpage

\clearpage

\clearpage
\begin{figure}
\centering
    \includegraphics[width=\textwidth]{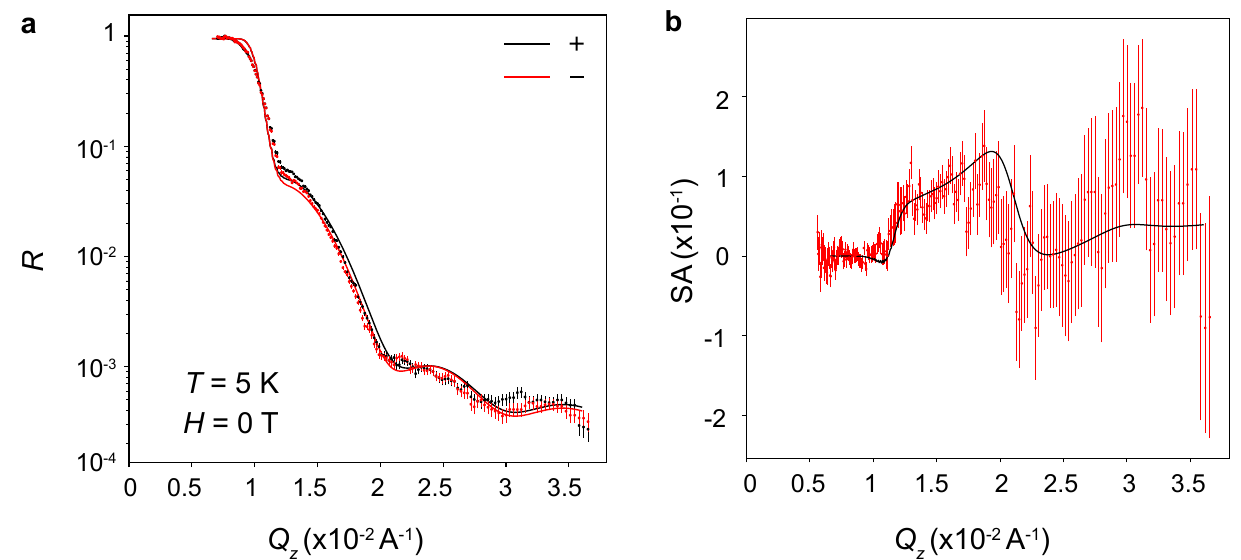}\vspace{-4mm}
    \caption{\textbf{Polarized neutron reflectivity measurement on the Nd honeycomb lattice at $\boldsymbol{T}$ = 4 K and $\boldsymbol{H}$ = 1 T}. \textbf{a} Fitted specular reflectivity of (+) and (-) channels. \textbf{b} the corresponding spin asymmetry calculated as ${\textrm{SA}} = \frac{(+) - (-)}{(+) + (-)}$. The spin asymmetry clearly indicates that the Nd honeycomb lattice already develops net magnetization under 1 T in-plane magnetic field application at $T$ = 4 K. The fitting share the same structural model parameters as used for $H$ = 0 T and 4.5 T presented in Fig. 3.}
    \label{fig:pnr_4K_1T}
\end{figure}

\clearpage
\begin{figure}
\centering
    \includegraphics[width=\textwidth]{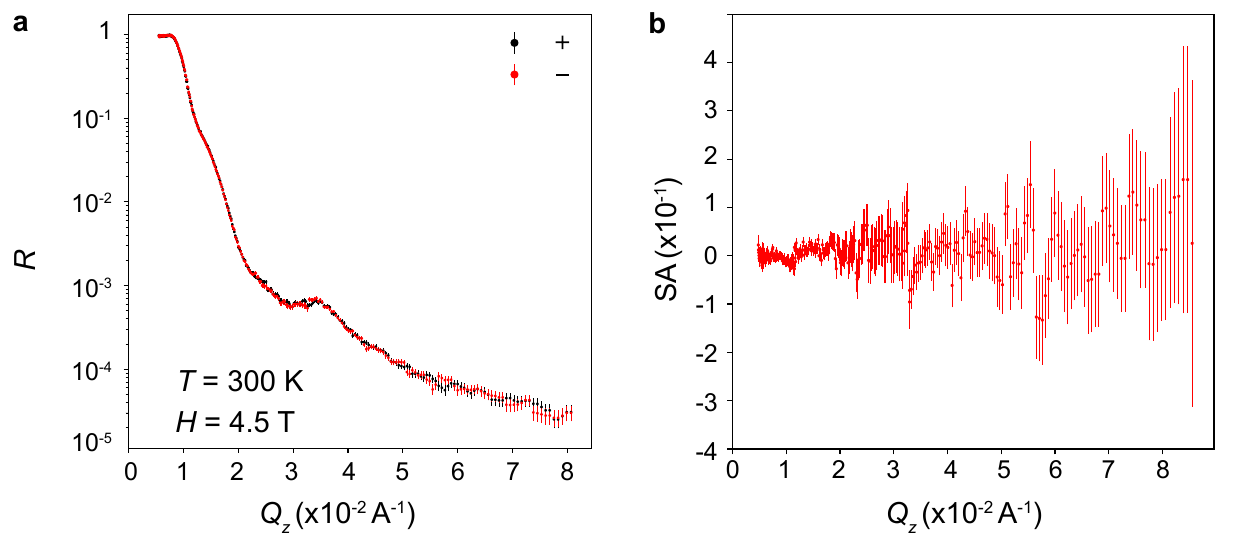}\vspace{-4mm}
    \caption{\textbf{Polarized neutron reflectivity measurement on the Nd honeycomb lattice at $\boldsymbol{T}$ = 300 K and $\boldsymbol{H}$ = 4.5 T.} \textbf{a} Fitted specular reflectivity of (+) and (-) channels. \textbf{b} the corresponding spin asymmetry calculated as ${\textrm{SA}} = \frac{(+) - (-)}{(+) + (-)}$. The overlapping of the (+) and (-) reflectivity curves suggests that no net magnetization is detected in the Nd honeycomb lattice at room temperature even at $H$ = 4.5 T in-plane magnetic field application.}
    \label{fig:pnr_300K_4p5T}
\end{figure}

\clearpage
\begin{figure}
\centering
    \includegraphics[width=\textwidth]{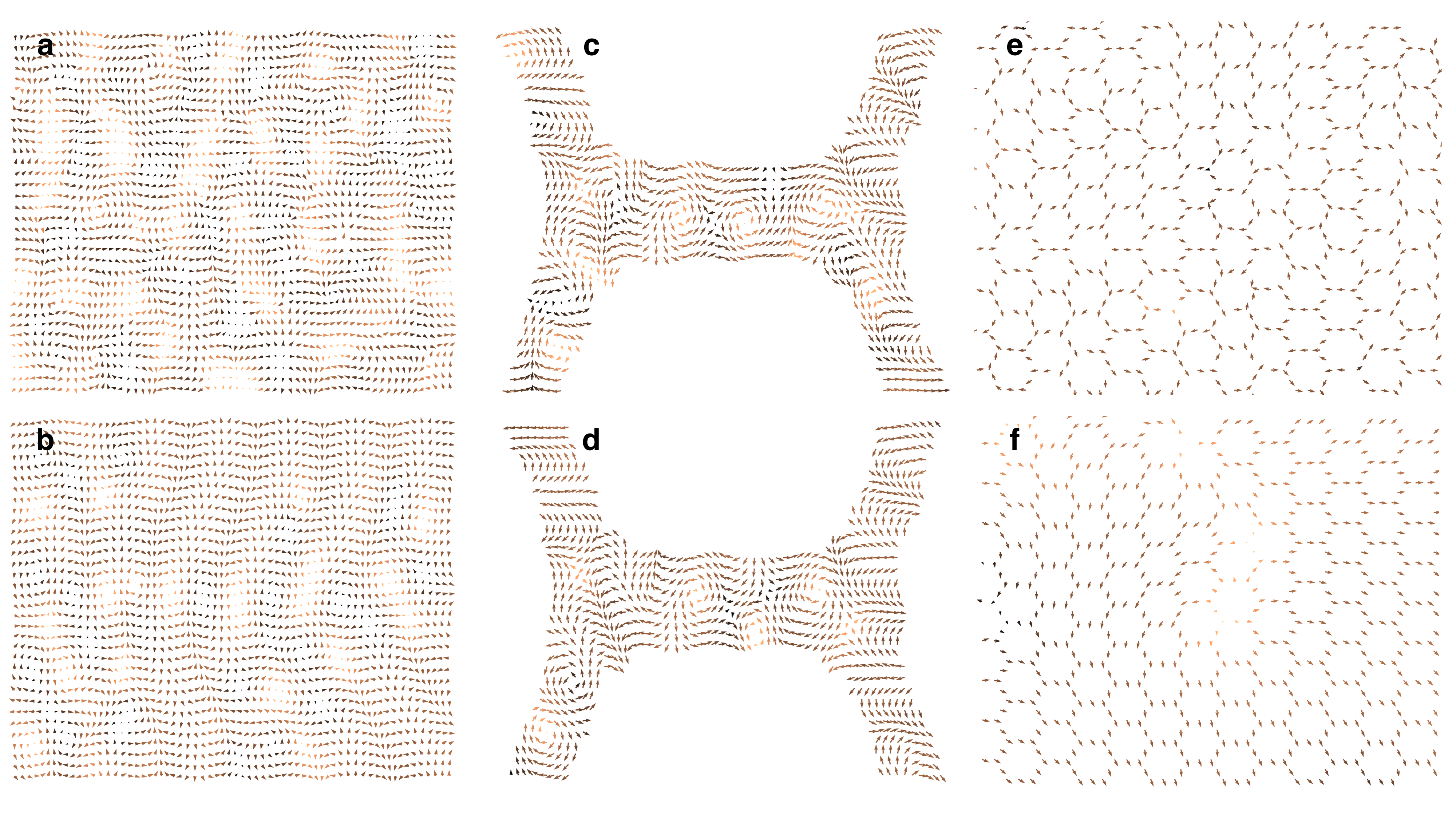}\vspace{-4mm}
    \caption{\textbf{Simulated Nd and Py magnetization.} Out of plane magnetization $M_z$ is indicated by color, with dark (light) coloration indicating negative (positive) $M_z$. \textbf{a} Simulated 6nm thin film Nd at 10ns after random initialization with a damping constant of 0.3. \textbf{b} Simulated 6nm thin film Nd at 100ns after random initialization with a damping constant of 0.3., illustrating the decay of quasiparticles and formation of locally ordered domains. \textbf{c} Simulated 6nm artificial honeycomb Nd at 10ns after random initialization with a damping constant of 0.3. \textbf{d} Simulated 6nm artificial honeycomb Nd at 100ns after random initialization with a damping constant of 0.3, illustrating the more robust nature of the quasiparticles in the artificial honeycomb case. \textbf{e} Simulated 6nm artificial honeycomb Py, with joint magnetizations predominantly oriented along their joint axis. \textbf{f} Simulated 9nm artificial honeycomb Py, showing joint magnetizations more freely orienting away from their joint axis.}
    \label{fig:theory}
\end{figure}